\magnification=1200
\settabs 18 \columns

\baselineskip=15 pt
\topinsert \vskip 0.75 in
\endinsert

\def\sqr#1#2{{\vcenter{\vbox{\hrule height.#2pt
 \hbox{\vrule width.#2pt height#1pt \kern#1pt
 \vrule width.#2pt} \hrule height.#2pt}}}}

\def\operp{\hbox{${\kern+.25em{\bigcirc}
\kern-.85em\bot\kern+.85em\kern-.25em}$}}

\def\lsim{\;\raise0.3ex\hbox{$<$\kern-0.75em\raise-1.1ex\hbox{$\sim$}}\;}
\def\gsim{\;\raise0.3ex\hbox{$>$\kern-0.75em\raise-1.1ex\hbox{$\sim$}}\;}
\def\no{\noindent}

\def\ce{\centerline}
\def\ve{\vfill\eject}
\def\rdots{\mathinner{\mkern1mu\raise1pt\vbox{\kern7pt\hbox{.}}\mkern2mu
 \raise4pt\hbox{.}\mkern2mu\raise7pt\hbox{.}\mkern1mu}}

\def\e e{$e^+ e^-$ }



\ce{{\bf $q$-ELECTROWEAK (II)}}
\vskip.5cm

\ce{\it R. J. Finkelstein}
\vskip.3cm
\ce{Department of Physics and Astronomy}
\ce{University of California, Los Angeles, CA  90095-1547}
\vskip1.0cm

\no{\bf Abstract.}  A gauged $SU_q(2)$ theory is characterized by two dual
algebras, the first lying close to the Lie algebra of $SU(2)$ while
the second introduces new degrees of freedom that may be associated with
non-locality or solitonic structure.  The first and second algebras,
here called the external and internal algebras respectively, define two
sets of fields, also called external and internal.  The gauged external
fields agree with the Weinberg-Salam model at the level of the doublet
representation but differ at the level of the adjoint representation.
For example the $g$-factor of the charged $W$-boson differs in the two
models.  The gauged internal fields remain speculative but are analogous
to color fields.

\ve

\line{{\bf 1. Introduction.} \hfil}
\vskip.3cm

In an earlier note we discussed a modification of the Weinberg-Salam model
suggested by gauging $SU_q(2)_L \times U(1)$.$^1$  It is reasonable to do
this since $SU(2)$, unlike the Poincar\'e group, is a phenomenological
group, and $SU_q(2)$ may also be phenomenologically useful.

In taking this step one finds that the Lie algebra gets replaced by two dual
algebras: the first lying close to and approaching the original Lie algebra
in a correspondence limit $(q=1)$ while the second algebra is new and
introduces new degrees of freedom. 

We propose to study the replacement of the point-particle classical field
theory by a soliton field theory described in the two complementary ways that
correspond to the two dual algebras.  In the first (macroscopic) picture the
particles are regarded as point-like but subject to the first algebra.  In the
complementary (microscopic) picture the solitons are regarded as composed
of preons subject to the second (dual) algebra.  The first algebra is little
different from the $SU(2)$ Lie algebra and will be called the external algebra.
The second algebra is exotic and will be called the internal algebra since
it governs the dynamics of the constituent particles.

\vskip.5cm

\line{{\bf 2. Irreducible Representations of $SU_q(2)$.} \hfil}
\vskip.3cm

We shall first summarize the necessary information about $SU_q(2)$.

The two-dimensional representation of $SL_q(2)$ may be defined by
$$
T\epsilon T^t = T^t\epsilon T = \epsilon \eqno(2.1)
$$
\no where $t$ means transpose and
$$
\epsilon = \left(\matrix{0 & q_1^{1/2} \cr
-q^{1/2} & 0 \cr} \right) \qquad q_1 = q^{-1} \eqno(2.2)
$$
\no Set
$$
T = \left(\matrix{\alpha & \beta \cr \gamma & \delta \cr} \right)
\eqno(2.3)
$$
\no Then 
$$
\eqalign{\alpha\beta &= q\beta\alpha \cr
\delta\beta &= q_1\beta\delta \cr
\hfil \cr} \qquad
\eqalign{\alpha\gamma &= q\gamma\alpha \cr
\delta\gamma &= q_1\gamma\delta \cr
\hfil \cr} \qquad
\eqalign{&\alpha\delta-q\beta\gamma = 1 \cr
&\delta\alpha-q_1\beta\gamma = 1 \cr
&\beta\gamma = \gamma\beta \cr} \eqno(2.4)
$$
\no If $q=1$, Eqs. (2.4) are satisfied by complex numbers and $T$ is
defined over a continuum, but if $q\not= 1$, then $T$ is defined only
over this algebra--a non-commuting space.

A two-dimensional representation of $SU_q(2)$ may be obtained by going to a
matrix representation of (2.4) and setting$^2$
$$
\gamma = -q_1\bar\beta \qquad \delta = \bar\alpha \eqno(2.5)
$$
\no where the bar means Hermitian conjugate.  Then
$$
\eqalign{\alpha\beta &= q\beta\alpha \cr
\alpha\bar\beta &= q\bar\beta\alpha \cr} \qquad
\eqalign{&\alpha\bar\alpha + \beta\bar\beta = 1 \cr
&\bar\alpha\alpha + q_1^2\bar\beta\beta = 1 \cr} \qquad
\beta\bar\beta = \bar\beta\beta \eqno(A)
$$
\no and $T$ is unitary:
$$
\bar T = T^{-1} \eqno(2.6)
$$
\no If $q=1$, $(A)$ may be satisfied by complex numbers and $T$ is a
$SU(2)$ unitary-simplectic matrix.  If $q\not= 1$, there are no finite
representations of $(A)$ unless $q$ is a root of unity.  We shall assume
that $q$ is real and $q<1$.

The irreducible representations of $SU_q(2)$ are as follows:$^3$
$$
{\cal{D}}^j_{mm^\prime}(\alpha,\bar\alpha,\beta,\bar\beta) =
\Delta^j_{mm^\prime}\sum_{s,t}\bigg\langle\matrix{n_+\cr s\cr}\bigg\rangle_1
\bigg\langle\matrix{n_-\cr t\cr}\bigg\rangle_1q_1^{t(n_+-s+1)}
(-)^t\delta(s+t,n_+^\prime)\alpha^s\beta^{n_+-s}\bar\beta^t
\bar\alpha^{n_--t} \eqno(2.7)
$$
\no where
$$
\eqalign{n_\pm &= j\pm m \cr
n_\pm^\prime &= j\pm m^\prime \cr} \qquad
\bigg\langle\matrix{n\cr s\cr}\bigg\rangle_1 = {\langle n\rangle_1!\over
\langle s\rangle_1!\langle n-s\rangle_1!} \qquad
\langle n\rangle_1 = {q_1^{2n}-1\over q_1^2-1}  \eqno(2.8)
$$
$$
\Delta^j_{mm^\prime} = \biggl[{\langle n_+^\prime\rangle_1!
\langle n^\prime_-\rangle_1!\over \langle n_+\rangle_1!\langle n_-\rangle_1!}\biggr]^{1/2} \qquad q_1 = q^{-1}
$$

Here the arguments of (2.7) satisfy the $(A)$ algebra.  In the limit
$q = 1$ ${\cal{D}}^j_{mm^\prime}$ become the Wigner functions,
${\cal{D}}^j_{mm^\prime}(\alpha\beta\gamma)$, the irreducible representation
of $SU(2)$.  The orthogonality properties of the ${\cal{D}}^j_{mm^\prime}$
may be expressed as follows:$^2$
$$
h(\bar{\cal{D}}^j_{mn}{\cal{D}}^{j^\prime}_{m^\prime n^\prime}) =
\delta^{jj^\prime}\delta_{mm^\prime}\delta_{nn^\prime}
{q^{2n}\over[2j+1]_q} \eqno(2.9)
$$
\no where
$$
[x]_q = {q^x-q_1^x\over q-q_1} \eqno(2.10)
$$
\no Here $h$ is a linear operator introduced by Woronowicz having the property
that \break 
$h[\bar{\cal{D}}^{j^\prime}_{mn}{\cal{D}}^{j^\prime}_{m^\prime n^\prime}]$ for $SU_q(2)$
corresponds to the integral over the group manifold of $SU(2)$.  The
coefficients describing the decomposition of the product of two irreducible
representations into the Clebsch-Gordan series may be computed with the aid
of (2.9).$^4$
\vskip.5cm

\line{{\bf 3. The Dual Algebras.} \hfil}
\vskip.3cm

The dual algebras may be exhibited in the following way.  The two-dimensional
representation may be Borel factored:
$$
{\cal{D}}^{1/2}(\alpha,\bar\alpha,\beta\bar\beta) = e^{B\sigma_+}
e^{\lambda\theta\sigma_3} e^{C\sigma_-} \eqno(3.1)
$$
\no The algebra $(A)$ of $(\alpha,\beta,\bar\alpha,\bar\beta)$ is then
inherited by $(B,C,\theta)$ as
$$
\eqalignno{&(B,C) = 0 \quad (\theta,B) = B \quad (\theta,C) = C & (3.2)\cr
&\lambda = \ln q & (3.3)\cr} 
$$
\no The $2j+1$ dimensional irreducible representation of $SU_q(2)$ shown in
(2.7) may be rewritten in terms of $(B,C,\theta)$.  Then by expanding to
terms linear in $(B,C,\theta)$ one has
$$
{\cal{D}}^j_{mm^\prime}(B,C,\theta) = {\cal{D}}^j_{mm^\prime}(0,0,0) +
B(J^j_B)_{mm^\prime} + C(J^j_C)_{mm^\prime} + 2\lambda\theta
(J^j_\theta)_{mm^\prime} + \ldots \eqno(3.4)
$$
\no where the non-vanishing matrix coefficients $(J_B^j)_{mm^\prime}$,
$(J^j_C)_{mm^\prime}$, and $(J^j_\theta)_{mm^\prime}$ are
$$
\eqalignno{\langle m-1|J_B^j|m\rangle &= (\langle j+m\rangle_1
\langle j-m+1\rangle_1)^{1/2} & (3.5B) \cr
\langle m+1|J_C^j|m\rangle &= 
(\langle j-m\rangle_1\langle j+m+1\rangle_1)^{1/2} & (3.5C) \cr
\langle m|J_\theta^j|m\rangle &= m & (3.5\theta) \cr}
$$
\no and where $\langle n\rangle_1$, as given by (2.8),
is a basic integer corresponding to $n$.  The $(B,C,\theta)$ and
$(J_B,J_C,J_\theta)$ are generators of two dual algebras satisfying the
following commutation rules:
$$
(J_B,J_\theta) = -J_B \quad (J_C,J_\theta) = J_C \quad
(J_B,J_C) = q_1^{2J-1}[2J_\theta] \eqno(3.6)
$$
$$
(B,C) = 0 \qquad (\theta,B) = B \qquad (\theta,C) = C \eqno(3.7)
$$
\no Here $[x]$ is given by (2.10).  The commutation relations (3.6)
follow from (3.5).

In the fundamental and adjoint representations of the external algebra the
commutation relations (3.6) simplify as follows:$^1$
\item{(a)} $J = 1/2$ \qquad (fundamental)
$$
(J_B,J_\theta) = -J_B \quad (J_C,J_\theta) = J_C \quad
(J_B,J_C) = 2J_\theta \eqno(3.8)
$$
\item{(b)} $J = 1$ \qquad (adjoint)
$$
(J_B,J_\theta) = -J_B \quad (J_C,J_\theta) = J_C \quad
(J_B,J_C) = \langle 2\rangle_1 J_\theta \eqno(3.9)
$$

The right-hand side of (3.6) is not linear in the $J_\theta$ generators unless
$J=1/2$ or $J=1$, and only in these cases do we speak of a ``Lie algebra"
or structure constants.

For these two cases let us introduce Hermitian generators as follows:
$$
\eqalign{ J_B &= J_1 + i~J_2 \cr
J_C &= J_1-i~J_2 \cr
J_\theta &= J_3 \cr} \eqno(3.11)
$$
\no Then
$$
\eqalignno{J = 1/2: \qquad &(J_m,J_n) = i~\epsilon_{mnp} J_p & (3.12) \cr
J = 1:~~ \qquad &(J_1,J_2) = i~{\langle 2\rangle_1\over 2} J_3 \cr
&(J_2,J_3) = i~J_1 & (3.13) \cr
&(J_3,J_1) = i~J_2 \cr}
$$
\no For the fundamental and the adjoint representations we may write
$$
\eqalignno{(J_a,J_b) &= f_{ab}^{~~m}J_m & (3.14) \cr
g_{ab} &= {\rm Tr}~ J_aJ_b & (3.15) \cr}
$$
\no where $f_{ab}^{~~m}$ and $g_{ab}$ correspond to the usual structure
constants and group metric and where
$$
\eqalignno{ &f_{abc} = f_{ab}^{~~m}g_{mc} & (3.16) \cr
\noalign{\hbox{If $J=1/2$}}
&f_{abc} = i~\epsilon_{abc} & (3.17) \cr
\noalign{\hbox{~~~$J=1$}}
&f_{abc} = i~\langle 2\rangle_1\epsilon_{abc} & (3.18) \cr}
$$
\no In both cases $f_{abc}$ has the important property of being completely
antisymmetric.$^1$

The metric $g_{ab} \sim\delta_{ab}$ in the fundamental representation but
in the adjoint representation
$$
g_{ab} = g_a\delta_{ab} \eqno(3.19)
$$
\no where 
$$
g_1=g_2=\langle 2\rangle_1 \quad{\hbox{and}} \quad g_3=2 \eqno(3.20)
$$
\vskip.5cm

\line{{\bf 4. The Internal Algebra and the Microscopic Picture.} \hfil}
\vskip.3cm

Let us expand a generic field operator in the irreducible representations
(2.7) as follows:
$$
\psi(x,\{\alpha\}) = \sum_{jmn} \varphi^j_{mn}(x) {\cal{D}}^j_{mn}(\{\alpha\}) \eqno(4.1)
$$
\no where $\{\alpha\}$ is an abbreviation for $\{\alpha\bar\alpha\beta\bar\beta\}$ and where
${\cal{D}}^j_{mn}$, and therefore $\psi(x)$, lies in the algebra. Here $\varphi^j_{mn}(x)$ is expanded in Fock annihilation and creation
operators:
$$
\varphi^j_{mn}(x) = {1\over (2\pi)^{3/2}} \int
{d\vec p\over (2p_o)^{1/2}}
\bigl[e^{-ipx}a_{mn}^j(\vec p) + e^{ipx}\bar a^j_{mn}(\vec p)\bigr]
\eqno(4.2)
$$
\no where the Lorentz tensor indices have been suppressed.

Under a gauge transformation $(T)$
$$
T\psi = \sum \varphi^j_{mn}(x) T{\cal{D}}^j_{mn}(\{\alpha\}) \eqno(4.3)
$$
\no where $T{\cal{D}}^j_{mn}$ still lies in the internal algebra.  Then
$$
T{\cal{D}}^j_{mn} = \sum \langle jmn|T|j^\prime m^\prime n^\prime\rangle 
{\cal{D}}^{j^\prime}_{m^\prime n^\prime}
$$
$$
T\psi = \sum \varphi^j_{mn}\langle jmn|T|j^\prime m^\prime n^\prime\rangle 
{\cal{D}}^{j^\prime}_{m^\prime n^\prime} \eqno(4.4)
$$
\no where
$$
{q^{2n^\prime}\over [2j^\prime+1]}\langle j^\prime m^\prime n^\prime|T|jmn\rangle
= h(\bar {\cal{D}}^{j^\prime}_{m^\prime n^\prime}T{\cal{D}}^j_{mn}) \eqno(4.5)
$$
\no and $h$ is the linear operator appearing in (2.9).  If, for example,
$T = {\cal{D}}^{j^{\prime\prime}}_{m^{\prime\prime}n^{\prime\prime}}$
then $\langle j^\prime m^\prime n^\prime|T|jmn\rangle =
h(\bar{\cal{D}}^{j^\prime}_{m^{\prime\prime} n^{\prime\prime}} {\cal{D}}^{j^{\prime\prime}}_{m^{\prime\prime}
n^{\prime\prime}}{\cal{D}}^j_{mn})$ 
is a $q$-Clebsch-Gordan coefficient.$^4$

The field quanta associated with $\varphi^j_{mn}(x)$ or ${\cal{D}}^j_{mn}(\{\alpha\})$ may be interpreted as composite particles
while the constituent or preon fields may be associated with the generators
of the internal algebra.  The total field operator may be interpreted as an
expansion in solitons, as we shall now show.$^5$

Let us illustrate with the global Hamiltonian of a scalar field.  We assume
$$
H^q = {1\over 2} h\int~:~\sum^3_0 \partial_k\bar\psi\partial_k\psi +
m_o^2\bar\psi\psi~:~d\vec x \eqno(4.6)
$$
\no Here $h$, defined in (2.9), is an average over the algebra.

At the global level $H^q$ is invariant under gauge transformations since $T$ is
unitary and therefore
$$
\eqalign{&(\bar\psi\psi)^\prime = (\bar\psi\bar TT\psi) = \bar\psi\psi \cr
&(\partial_k\bar\psi\partial_k\psi)^\prime = \partial_k\bar\psi\partial_k\psi \cr} \eqno(4.7)
$$
\no By (4.1) and (4.2)
$$
H^q = \int d\vec p~ p_o \sum_{\scriptstyle jmn\atop\scriptstyle 
j^\prime m^\prime n^\prime} h(\bar{\cal{D}}^j_{mn}{\cal{D}}^\prime_{m^\prime n^\prime}){1\over 2}:\bigl[\bar a^j_{mn}(p)a^{j^\prime}_{m^\prime n^\prime}(p)
+ a^{j^\prime}_{m^\prime n^\prime}(p)a^j_{mn}(p)\bigr] \eqno(4.8)
$$
\no Evaluate $H^q$ on the state $|N(p);jmn\rangle$.  Then by (2.9)
$$
H^q|N(\vec p);jmn\rangle = \sum_{jmn} {Np_oq^{2n}\over [2j+1]_q}
|N(\vec p);jmn\rangle \eqno(4.9)
$$
\no Therefore the rest mass of a single field quantum associated with the
field $\varphi^j_{mn}$ is
$$
{m_oq^{2n}\over [2j+1]_q} \eqno(4.10)
$$
\no If $q=1$ the rest mass of a particle with quantum numbers $(jmn)$ is
$$
{m_o\over 2j+1}
$$
\no and does not depend on $n$.  If $q\not= 1$ the mass depends on $n$ and if
$q\cong 1$ there is an approximate harmonic oscillator fine structure.
On the other hand a point particle has no mass spectrum and the existence
of such a spectrum here implies an extended object.  Since the spectrum is
approximately that of a $q$-harmonic oscillator, one may assume that the
extension is approximately described by a $q$-harmonic oscillator wave
function.  It is in this sense that we describe the field quanta of $\psi$
as solitons.
\vskip.5cm

\line{{\bf 5. The External Algebra and $q$-Electroweak.} \hfil}
\vskip.3cm

In the Weinberg-Salam model the Lagrangian density is$^6$
$$
\eqalign{{\cal{L}} = &- {1\over 4}(G^{\mu\nu}\cdot G_{\mu\nu} + H^{\mu\nu}H_{\mu\nu}) +i(\bar LD\!\!\!/L + \bar RD\!\!\!/R) \cr
&+(\overline{D\phi})\cdot (D\phi)-V(\bar\varphi\varphi) - {m\over \rho_o}
(\bar L\varphi R + \bar R\bar\varphi L) \cr} \eqno(5.1)
$$
\no where the covariant derivative is
$$
D = \partial + ig\vec W\vec t + ig^\prime W_ot_o \eqno(5.2)
$$
\no Here $\vec W^\mu$ and $W_o^\mu$ are the connection fields of
$SU(2)_L$ and $U(1)$, the chiral isotopic spin and hypercharge groups with
independent coupling constants $g$ and $g^\prime$, while $G$ and $H$ are the
corresponding field strengths.  The Lagrangian (5.1) also contains the
contribution of one lepton doublet and the mass generating Hibbs doublet
$\varphi$.

In (5.2), the expression for the covariant derivative, the matrices $\vec t$
and $t_o$ are the generators of the $SU(2)$ and $U(1)$ groups.  If we now
pass to $SU_q(2)$ without changing $U(1)$, Eqs. (5.2) will be unchanged
in the doublet representation since
Eqs. (4.1) and (4.2) hold for both $SU(2)$ and $SU_q(2)$.  Therefore at the
level of the doublet representation there is no divergence between the
standard $SU(2)$ theory and the corresponding $SU_q(2)$ theory and one again
obtains the standard relations
$$
\eqalignno{e &= g\sin\theta_W = g^\prime\cos\theta_W & (5.3) \cr
M_W &= M_{\bf Z}\cos\theta_W & (5.4) \cr}
$$
\no where $g$ and $g^\prime$ are the coupling constants of the chiral
isotopic spin group and the hypercharge group respectively while $M_W$ and
$M_{\bf Z}$, the masses of the charged and neutral bosons, are also
related by $\theta_W$ the Weinberg angle.  The argument leading to these results is not
changed since the form of $D$ in (5.2) is not changed on interpreting the
$\vec t$ matrices as belonging to the fundamental representation of 
$SU_q(2)$ instead of $SU(2)$.

\vskip.5cm

\line{{\bf 6. Gauge Invariance of the External Sector.} \hfil}
\vskip.3cm

All matrices and fields in the external sector are numerically valued.
Consider a general field transformation:
$$
\psi^\prime = T\psi \eqno(6.1)
$$
\no By definition the covariant derivative, $D_\mu\psi$, then transforms as
follows:
$$
\eqalignno{(D_\mu\psi)^\prime &= T(D_\mu\psi) \cr
\noalign{\hbox{Hence}}
D_\mu^\prime &= TD_\mu T^{-1} & (6.2) \cr}
$$
\no In terms of $D_\mu$ the vector connection $W_\mu$ and the field strength
$G_{\mu\lambda}$ are defined by
$$
\eqalignno{W_\mu &= D_\mu-\partial_\mu & (6.3) \cr
G_{\mu\lambda} &= (D_\mu,D_\lambda) & (6.4) \cr}
$$
\no Then
$$
\eqalignno{W_\mu^\prime &= TW_\mu T^{-1} + T\partial_\mu T^{-1} & (6.5) \cr
G_{\mu\lambda}^\prime &= TG_{\mu\lambda} T^{-1} & (6.6) \cr}
$$
The field invariant may be chosen to be
$$
I = {\rm Tr}~G^{\mu\lambda}G_{\mu\lambda} \eqno(6.7)
$$
\no Assume now
$$
W_\mu = ig~W_\mu^at_a \eqno(6.8)
$$
\no where the numerically valued $t_a$ belong to the adjoint representation and satisfy equations
of the form (3.14).  Then
$$
G_{\mu\lambda} = G_{\mu\lambda}^a t_a \eqno(6.9)
$$
\no By (6.7) and (6.9)
$$
\eqalign{I &= {\rm Tr}~G_{\mu\lambda}^a G^{b\mu\lambda}t_at_b \cr
&= g_{ab} G^a_{\mu\lambda} G^{b\mu\lambda} \cr} \eqno(6.10)
$$
\no In the Weinberg-Salam model
$$
g_{ab} = 2\delta_{ab} \eqno(6.11)
$$
\no In that case one may write
$$
I = 2G^a_{\mu\lambda}G^{a\mu\lambda} = 2G_{\mu\lambda}\cdot
G^{\mu\lambda} \eqno(6.12)
$$
\no as in (5.1).  Here however we must retain $g_{ab}$ since by (3.20) it is
not isotropic.  Other terms like $\bar LDL$ are also invariant since $T$
is unitary and therefore
$$
\bar L^\prime D^\prime L^\prime = (\bar LT^{-1})(TDT^{-1})(TL) =
\bar LDL \eqno(6.13)
$$
\no Hence the full (5.1) is gauge-invariant with the external $SU_q(2)_L$
substituted for $SU(2)_L$.  As already remarked this $q$-theory leads to the
same consequences as the Weinberg-Salam theory at the doublet level. 
However there will be differences at the adjoint level.  In particular the
couplings of $W^+_\mu$ and $W^-_\mu$ appearing in $D_\mu$ depend on $q$
but those of $A_\mu$ and $Z_\mu$ do not.  Since the full theory remains
gauge invariant, however, there will still be Ward identities.

The deviations from the standard theory can also be seen by examining the
self-interactions of the vector fields, namely
$$
-{1\over 4g^2} g_{mn} G^m_{\mu\nu}G^{n\mu\nu} \eqno(6.14)
$$
\no where
$$
G^a_{\mu\nu} = \partial_\mu W^a_\nu-\partial_\nu W^a_\mu
 + f^q_{~bc} W^b_\mu W^c_\nu \eqno(6.15)
$$
\no and $g$ is the weak coupling constant appearing in (6.8).  The trilinear couplings are then
$$
\eqalign{\sim &g_{mn}f^m_{bc}W^b_\mu W^c_\nu(\partial^\mu W^{\nu n}
-\partial^\nu W^{\mu n}) \cr
&= f_{nbc}W^b_\mu W^c_\nu(\partial^\mu W^{\nu n}-\partial^\nu W^{\mu n}) \cr}
\eqno(6.16)
$$
\no where
$$
\eqalign{f_{nbc} &= i\langle 2\rangle_1\epsilon_{nbc} \quad
{\rm by ~ (3.18)} \cr
&= i(1+q_1^2)\epsilon_{nbc} \cr} \eqno(6.17)
$$
\no Hence the asymmetry expressed by $f_{bc}^{~m}$ may be removed in these
terms.

The quartic couplings are on the other hand
$$
\sim g_{mn} f^m_{bc} f^n_{k\ell} W^b_\mu W^c_\nu W^{k\mu}W^{\ell\nu}
\eqno(6.18)
$$
\no Here
$$
g_{mn}f^m_{bc}f^n_{k\ell} = f_{nbc}f^n_{k\ell} \eqno(6.19)
$$
\no At this point the asymmetry arising from (3.19) and expressed by
$f_{k\ell}^{~n}$ can no longer be hidden.  It distinguishes one preferred direction in
isotopic spin space, and in principle should be experimentally detectable.

There is in fact already a theoretically detectable divergence from the
Weinberg-Salam theory buried in the trilinear terms.  By (6.16) and (6.17)  
the trilinear terms are
$$
(1+q_1^2)g~\epsilon_{nbc}W^b_\mu W^c_\nu(\partial^\mu W^{\nu n}-
\partial^\nu W^{\mu n}) \eqno(6.20)
$$
\no The electromagnetic part of this interaction contains the term
$$
-ie(1+q_1^2) W_+^\mu W_-^\nu F_{\mu\nu} \eqno(6.21)
$$
\no which is obtained by use of the Weinberg-Salam relations:
$$
\eqalignno{W_\mu^3 &= A_\mu\sin\theta & (6.22a) \cr
\noalign{\hbox{and}}
e &= g\sin\theta & (6.22b) \cr}
$$
\no These relations also hold here since their derivation is at the level of
the doublet representation.

A general term of this kind, namely
$$
-ie\kappa W_+^\mu W_-^\nu F_{\mu\nu} \eqno(6.23)
$$
\no gives rise to a magnetic moment
$$
(1+\kappa){e\over 2m_W} \vec s \eqno(6.24)
$$
\no where $\vec s$ is the spin vector.

The $g$-factor of the $W$ boson in the $q$-model is then $1+q_1^2$ rather than
$g=2$, the value in the Weinberg-Salam model.$^6$

It is remarkable that the $gGG$ in (6.14) remains gauge invariant although both
the cubic and quartic terms are changed.  Since gauge invariance is still
preserved with the new structure constants, the good formal
properties of the standard theory are also preserved.  Except for the appearance
of structure constants depending on $q$ the field Lagrangian is standard.

In the Weinberg-Salam theory the component fields $(W_1,W_2,W_3)$ appear in
two ways: first, in interaction with chiral fermions the three fields are
associated with matrix elements of the two-dimensional fundamental
matrices; and second, in the description of free fields the $W_i$ are associated with the
three-dimensional generators $t_i$ rather than with the matrix elements of the
$t_i$.  We have followed the same pattern here, but there remains a
difference between the fundamental and adjoint representation in the two
formalisms.  In the Standard Model these two representations are related
by the Clebsch-Gordan coefficients of $SU(2)$.  Here  on the other hand${\cal{D}}^{1/2}(\alpha,\bar\alpha,\beta\bar\beta)$ and ${\cal{D}}^1(\alpha,\bar\alpha,\beta,\bar\beta)$ 
in the internal algebra are related by the
Clebsch-Gordan coefficients computed from the algebra $(A)$.$^4$  Next one may go from the fundamental to the adjoint representation of the
external algebra indirectly by going through the internal algebra
and then making use of (3.4).  Alternatively one may make use of the
co-product defined by:$^7$
$$
\eqalign{\Delta(J^{1/2}_z)&\buildrel \rm def \over = 1\otimes J_z^{1/2} +
J_z^{1/2} \otimes 1 \cr
\Delta(J^{1/2}_\pm) &\buildrel \rm def \over = \hat q^{-{1\over 2} J_z}
\otimes J^{1/2}_\pm + J^{1/2}_\pm \otimes \hat q^{{1\over 2} J_z} \cr}
\eqno(6.25)
$$
\no and decomposing the 4-dimensional representation so obtained into the
adjoint and trivial representations.  The two procedures are equivalent.

Here note the following relation between $\hat q$ and $q$:
$$
[2]_{\hat q} = \langle 2\rangle_1 \eqno(6.26)
$$
\no This shift from $\hat q$ to $q$ results from our use of $\langle~~\rangle_1$
in (3.5) instead of $[~~]_q$.

In spite of these differences between the external $q$-algebra and the Lie
algebra, the external physical fields, as well as their Lagrangian and transformation laws, are all numerically valued and can be treated by the
standard procedures that we have followed here.
\vskip.5cm

\line{{\bf 7. Gauge Invariance of the Internal Sector.} \hfil}
\vskip.3cm

To seriously pursue the $q$-theory one must discuss the dual algebra
generated by $(B,C,\theta)$ or alternatively by $(\alpha,\bar\alpha,\beta,\bar\beta)$.  Since this algebra is not a Lie
algebra, any gauge theory based on the dual algbra must be quite different
from the gauge theory based on $(J_B,J_C,J_\theta)$.  In particular there is
no analogue of $g_{ab}$.

Nevertheless one may still define a vector connection $V_\mu$ in terms of the
covariant derivative $\nabla_\mu$ as before:
$$
\nabla_\mu = \partial_\mu + V_\mu \eqno(7.1)
$$
\no and the corresponding field strengths:
$$
V_{\mu\nu} = (\nabla_\mu,\nabla_\nu) \eqno(7.2)
$$
\no The earlier stated transformation laws (6.2) and (6.6) still hold
$$
\eqalignno{\nabla_\mu^\prime &= T\nabla_\mu T^{-1} & (7.3) \cr
V_{\mu\nu}^\prime &= TV_{\mu\nu} T^{-1} & (7.4) \cr}
$$
\no where $T$ lies in the internal algebra $A$.  Then
$$
(V_{\mu\nu} V^{\mu\nu})^\prime = T(V_{\mu\nu}V^{\mu\nu})T^{-1} \eqno(7.5)
$$
\no In the standard theory the field invariant may be expressed as either
${\rm Tr}~G^{\mu\nu}G_{\mu\nu}$ or $g_{ab}G^a_{\mu\nu}G^{b\mu\nu}$.  That
is not possible here because both $V^{\mu\nu}$ and $T$ lie in the $A$-algebra.
The trace is therefore not invariant since in general
$$
((V^{\mu\nu}){ab},T_{cd}) \not= 0
$$

Therefore we choose as field invariant
$$
I = h(\Phi^+ V_{\mu\lambda} V^{\mu\lambda}\Phi) \eqno(7.6)
$$
\no where
$$
\eqalignno{\Phi^\prime &= T\Phi & (7.7) \cr
\Phi^{+\prime} &= \Phi^+T^{-1} & (7.8) \cr}
$$
\no and $T$ is a unitary transformation lying in the $A$-algebra.

The new factor $\Phi$ may be taken to be a Higgs field.  Then
$$
\eqalign{I &= \sum h\bigl[\varphi^{+j}_{mn}({\cal{D}}^j_{mn})^+ \cdot
(V_{\mu\lambda})^{j^\prime}_{m^\prime n^\prime}{\cal{D}}^{j^\prime}_{m^\prime n^\prime} \cdot(V^{\mu\lambda})^{j^{\prime\prime}}_{m^{\prime\prime} n^{\prime\prime}}{\cal{D}}^{j^{\prime\prime}}_{m^{\prime\prime} n^{\prime\prime}}\cdot \varphi^{j^{\prime\prime\prime}}_{m^{\prime\prime\prime}
n^{\prime\prime\prime}}{\cal{D}}^{j^{\prime\prime\prime}}_{m^{\prime\prime\prime}n^{\prime\prime\prime}}\bigr] \cr
&= \sum(\varphi^+)^j_{mn}(V_{\mu\lambda})^{j^\prime}_{m^\prime n^\prime}
(V^{\mu\lambda})^{j^{\prime\prime}}_{m^{\prime\prime}n^{\prime\prime}}
\varphi^{j^{\prime\prime\prime}}_{m^{\prime\prime\prime}n^{\prime\prime\prime}}
h[({\cal{D}}^j_{mn})^+{\cal{D}}^{j^\prime}_{m^\prime n^\prime}
{\cal{D}}^{j^{\prime\prime}}_{m^{\prime\prime} n^{\prime\prime}}
{\cal{D}}^{j^{\prime\prime\prime}}_{m^{\prime\prime\prime}
n^{\prime\prime\prime}}] \cr} \eqno(7.9)
$$
\no In particular if $I$ is evaluated on the vacuum state of $\Phi$ one
finds
$$
I = [(\varphi^+)^o_{oo}\varphi^o_{oo}]\sum(V_{\mu\lambda})^{j^\prime}_{m^\prime n^\prime} (V^{\mu\lambda})^{j^{\prime\prime}}_{m^{\prime\prime}n^{\prime\prime}}
\biggl[\delta^{j^\prime j^{\prime\prime}}\delta_{m^\prime m^{\prime\prime}}
\delta_{n^\prime n^{\prime\prime}}{1\over [2j^\prime+1]_q}\biggr] \eqno(7.10)
$$
\no or
$$
I = (\varphi^o_{oo})^2 \sum_{jmn}(V_{\mu\lambda})^j_{mn}
(V^{\mu\lambda})^j_{mn} {q^{2n}\over [2j+1]_q} \eqno(7.11)
$$
\no The contribution of the trivial representation to this sum is
$$
I_o = |\varphi^o_{oo}|^2(V_{\mu\lambda})^o_{oo}(V^{\mu\lambda})^o_{oo}
\eqno(7.12)
$$
\no which resembles
$$
{1\over g^2} V_{\mu\lambda}V^{\mu\lambda} \eqno(7.13)
$$
\no for the Abelian case if we set
$$
(\varphi^o_{oo})^2 = {1\over g^2} \eqno(7.14)
$$
\no The full $I$ contains contributions that are averaged over all
representations.  

One may choose the following gauge invariant
Lagrangian$^8$
$$
h\int\biggl\{-{1\over 4} \Phi^+V_{\mu\lambda}V^{\mu\lambda}\Phi +
i~\bar\psi\gamma^\mu\nabla_\mu\psi + {1\over 2}
[\nabla_\mu\Phi^+\nabla^\mu\Phi] + U(\Phi^+\Phi)\biggr\}
\eqno(7.15)
$$

This form differs from (4.2) in Ref. 8 in two respects: (a) it is an average
over the algebra and (b) the invariance transformations are unitary.

\vskip.5cm

\line{{\bf 8. Macroscopic and Microscopic Pictures.} \hfil}
\vskip.3cm

There are two possibilities suggested by the foregoing.  Following the first
or standard path one may expand the external gauge fields in the numerical matrices
$(J_1,J_2,J_3)$ as well as in the usual normal modes.
Following the second path one may expand the internal fields in the
${\cal{D}}^j_{mn}(\alpha,\bar\alpha,\beta,\bar\beta)$ as well as in the usual
normal modes.  In the first case one has the standard classical point
particle theory obeying the algebra (3.12).   In the second case one has a
classical soliton field theory lying in the algebra of the arguments of
${\cal{D}}^j_{mn}(\alpha,\bar\alpha,\beta,\bar\beta)$ as illustrated in
(4.6) and the following discussion.  The point particle picture is
macroscopic while the soliton picture is microscopic.  Analogous to the
treatment of familiar composite particles by the separation into relative and
center of mass coordinates, or into an internal and an external problem, we
try to give one (internal) description based on the ${\cal{D}}^j_{mn}$
which might be called the color description and a second (external)
description based on the algebra of $(J_1,J_2,J_3)$ which might be called
the flavor description.

To pass from the field operator in the microscopic description to the
corresponding operator in the macroscopic description one averages the
operator field of the soliton over the algebra as follows:
$$
\eqalignno{h[\psi(x)] &= h\bigl[\sum\varphi^j_{mn}(x){\cal{D}}^j_{mn}\bigr]
& (8.1) \cr
&= \sum \varphi^j_{mn}(x) h({\cal{D}}^j_{mn}) \cr
&= \varphi^o_{oo}(x) & (8.2) \cr}
$$
\no We interpret $\varphi^o_{oo}(x)$ as the field operator in the point
particle description.

In the standard quantum field theory the field quanta acquire mass and
extension via clouds of virtual particles or renormalization of the bare mass.
Solitons also arise in classical theory, both topologically and
non-topologically, in many forms including the Prasad-Sommerfield model and in the context of Kaluza-Klein extensions, as strings and branes.  The proposal
described here offers another classical point of departure for a modification of quantum field theory.  Finally we emphasize that the model here proposed is suggested by and derived from the quantum group $SU_q(2)$ but does not strictly adhere to the structure of the quantum group, as ordinarily understood.
\vskip.5cm

\line{{\bf Acknowledgements.} \hfil}
\vskip.3cm

I thank E. D'Hoker for useful discusssion.

\vskip.5cm

\line{{\bf References.} \hfil}
\vskip.3cm

\item{1.} R. Finkelstein, hep-th/010075.
\item{2.} S. Woronowicz, Commun. Math. Phys. {\bf 111}, 613 (1987).
\item{3.} R. Finkelstein, Lett. Math. Phys. {\bf 29}, 75 (1993); hep-th/0106283.
\item{4.} See for example, C. Cadavid and R. Finkelstein, J. Math. Phys.
{\bf 36}, 1912 (1995).
\item{5.} R. Finkelstein, hep-th/0106283.
\item{6.} See for example, K. Huang, {\it Quarks, Leptons and Gauge Fields},
World Scientific (1982), p. 109.
\item{7.} See for example, Biedenharn and Lohe, {\it Quantum Groups, Symmetry
and $q$-Tensor Algebras}, World Scientific (1999), p. 17.
\item{8.} R. Finkelstein, Mod. Phys. Lett. A{\bf 15}, 1709 (2000).
\end
\bye